\documentclass[12pt]{article}   
\pdfoutput=1
\usepackage{epsf,color,colordvi}
\usepackage{graphicx}
\usepackage{amsmath}
\usepackage{verbatim}
\usepackage{amssymb}
\usepackage[most]{tcolorbox}
\usepackage{cite}
\usepackage{fontenc}
\usepackage{float}
\usepackage[lofdepth,lotdepth,caption=false]{subfig}
\usepackage{color}
\usepackage{booktabs}
\usepackage{tabularx}
\usepackage{multirow}
\usepackage{longtable}
\usepackage{lscape}
\usepackage[font=small,skip = 0pt]{caption}
\usepackage{dcolumn}
\usepackage{rotating}
\usepackage{xcolor}
\usepackage{colortbl}
\usepackage{hyperref}\usepackage[normalem]{ulem}
\definecolor{darkgreen}{cmyk}{1,0,1,0.4}
\definecolor{brown}{cmyk}{0,0.8,1,0.2}
\definecolor{darkred}{cmyk}{0,1,1,0.2}

\allowdisplaybreaks[4] 
\makeatletter
\renewcommand{\fnum@table}{\textbf{\tablename~\thetable}}
\renewcommand{\fnum@figure}{\textbf{\figurename~\thefigure}}
\makeatother

\newcounter{myenumi}

\renewcommand{\themyenumi}{\roman{myenumi}}
{\end{list}}

\newlength{\myem}
\newcounter{mysubequation}[equation]

\makeatletter
\renewcommand{\section}{\@startsection{section}{1}{0em}{-\baselineskip}%
{\baselineskip}{\normalfont\large\bfseries}}
\renewcommand{\subsection}%
{\@startsection{subsection}{2}{0em}{-0.7\baselineskip}%
{0.7\baselineskip}{\normalfont\bfseries}}
\makeatother

\newcommand{\bi}{\begin{itemize}}
\newcommand{\ei}{\end{itemize}}

\def\beq{\begin{equation}}
\def\eeq{\end{equation}}

\newcommand{\bea}{\begin{eqnarray}}
\newcommand{\eea}{\end{eqnarray}}

\newcommand{\td}{\theta_{14}}
\newcommand{\te}{\theta_{24}}
\newcommand{\tf}{\theta_{34}}
\newcommand{\da}{\delta_{13}}
\newcommand{\db}{\delta_{14}}

\newcommand{\ldm}{\Delta m_{31}^2}
\newcommand{\sdm}{\Delta m_{21}^2}
\newcommand{\lldm}{\Delta m_{41}^2}

\def\epsilon{\varepsilon}

\newcommand{\chisq}{\ensuremath{\chi^2}}

\def\<{\langle}
\def\>{\rangle}

\def\dfrac#1#2{{\displaystyle\frac{#1}{#2}}}

\def\lsim{\mathrel{\rlap{\lower4pt\hbox{\hskip1pt$\sim$}}
    \raise1pt\hbox{$<$}}}         
\def\gsim{\mathrel{\rlap{\lower4pt\hbox{\hskip1pt$\sim$}}
    \raise1pt\hbox{$>$}}}         

\newcommand{\dacp}[1]{\ensuremath{\delta [\Delta P^{CP/T}_{\alpha\beta}]}}

\newcommand{\pbarab}[1]{\ensuremath{{ P}_{\bar{\alpha} \bar{\beta}} }}
\newcommand{\pbarba}[1]{\ensuremath{{ P}_{\bar{\beta} \bar{\alpha}} }}
\newcommand{\acpab}[1]{\ensuremath{A^{CP}_{\alpha \beta}}}
\newcommand{\acpaa}[1]{\ensuremath{A^{CP}_{\alpha \alpha}}}
\newcommand{\ataa}[1]{\ensuremath{A^{T}_{\alpha \alpha}}}
\newcommand{\acpba}[1]{\ensuremath{A^{CP}_{\beta \alpha}}}
\newcommand{\atab}[1]{\ensuremath{{A}^{T}_{\alpha \beta}}}
\newcommand{\atba}[1]{\ensuremath{{A}^{T}_{\beta \alpha}}}
\newcommand{\acptab}[1]{\ensuremath{A^{CPT}_{\alpha \beta}}}

\newcommand{\dpcpme}[1]{\ensuremath{\Delta {P}^{CP(3+1)}_{\mu e} }}
\newcommand{\dpcpmt}[1]{\ensuremath{\Delta { P}^{CP}_{\mu \tau} }}
\newcommand{\dpcpet}[1]{\ensuremath{\Delta { P}^{CP}_{e \tau } }}
\newcommand{\dpcpee}[1]{\ensuremath{\Delta { P}^{CP}_{e e} }}
\newcommand{\dpcpmm}[1]{\ensuremath{\Delta { P}^{CP(3+1)}_{\mu \mu} }}
\newcommand{\dpcptt}[1]{\ensuremath{\Delta { P}^{CP}_{\tau \tau } }}
\newcommand{\dptme}[1]{\ensuremath{\Delta { P}^{T}_{\mu e}}}
\newcommand{\dptmt}[1]{\ensuremath{\Delta { P}^{T}_{\mu \tau}}}
\newcommand{\dptet}[1]{\ensuremath{\Delta { P}^{T}_{e \tau}}}
\newcommand{\dpcptab}[1]{\ensuremath{\Delta { P}^{CPT}_{\alpha \beta} }}

\begin{document}
\begin{titlepage}
\vspace*{-3.cm}
\renewcommand{\thefootnote}{\fnsymbol{footnote}}
\setcounter{footnote}{-1}

{\begin{center}
{\large\bf 
Impact of improved energy resolution on DUNE sensitivity 
in presence of a light sterile neutrino
\\[0.2cm]
}
\end{center}}

\renewcommand{\thefootnote}{\alph{footnote}}

\vspace*{.8cm}
\vspace*{.3cm}

{
\begin{center} 

   {\sf                 Sabila Parveen$^{\S}$\,  \footnote[1]{\makebox[1.cm]{Email:} sabila41\_sps@jnu.ac.in},
                }
        {\sf               Jogesh Rout$^{\P}$\,  \footnote[2]{\makebox[1.cm]{Email:} johesh.rout1@gmail.com}
        }
        and                
            {\sf                 Poonam Mehta$^{\S}$\,\footnote[3]{\makebox[1.cm]{Email:} pm@jnu.ac.in}
}
\end{center}
}
\vspace*{0cm}
{\it 
\begin{center}
$^\S$\, School of Physical Sciences, Jawaharlal Nehru University, 
      New Delhi 110067, India  \\
$^\P$\, Department of Physics, Shree Ram College, Rampur, Subarnapur, Odisha 767045, India   \\
\end{center}
}

{\Large 
\bf
 \begin{center} Abstract  
  
\end{center} 
 } 

The Deep Underground Neutrino Experiment (DUNE) primarily aims to measure the yet unknown parameters of the standard three neutrino framework, i.e., the determination of Dirac CP phase ($\delta_{13}$), neutrino mass hierarchy (MH) and octant of $\theta_{23}$. In the present work, we consider the standard three neutrino paradigm (referred to as the $(3+0)$ case) and beyond with an additional light sterile neutrino (referred to as the $(3+1)$ case). 
We consider two configurations : standard energy resolution as in DUNE Technical Design Report (TDR) and improved energy resolution and study the impact of energy resolution in the $(3+0)$ and $(3+1)$ cases. 
In general, inclusion of subdominant new physics effects spoils the sensitivities. However, improved energy resolution leads to enhancement in sensitivities to the three unknowns in both $(3+0)$ and $(3+1)$ cases.

\vspace*{.5cm}

\end{titlepage}

\newpage

\renewcommand{\thefootnote}{\arabic{footnote}}
\setcounter{footnote}{0}
\section{Introduction}

The immense progress over the past few decades in neutrino physics with various ongoing and upcoming efforts has opened up possibilities of probing physics beyond the Standard Model (BSM) with neutrinos. 
The wealth of data accumulated from a series of oscillation experiments using solar, atmospheric, accelerator and reactor neutrinos in past few decades have established beyond doubt that neutrinos oscillate among the three flavor while conserving the lepton number~\cite{nobel2015} (for global analyses of neutrino data, see~\cite{deSalas:2020pgw,Capozzi:2021fjo,Esteban:2024eli}). The neutrino oscillation experiments are sensitive only to the mass-squared differences (but not the absolute masses). There are several remaining questions at the present juncture and in the present work, we shall explore some of the yet unresolved questions such as establishing the presence (or absence) of CP violation in the leptonic sector, determining the neutrino MH and the correct octant of $\theta_{23}$. 

The question of whether CP in the leptonic sector is violated or not is a fundamental one and could have a bearing on matter-antimatter asymmetry of the universe~\cite{Davidson:2008bu}. 
The issue of determining the hierarchy of neutrino masses $m_i$, with $i = 1, 2, 3$, refers to finding the ordering of the mass eigenstates. The present oscillation data (i.e., the smaller mass-squared difference $\Delta m^2_{21} = m_2^2 - m_1 ^2 > 0$  and the absolute value of the larger mass-squared splitting, $|\Delta m^2_{31}| = |m_3^2 - m_1^2|$) indicates a hierarchical pattern of neutrino masses and one can have the following two possibilities : normal hierarchy (NH) when sign of $\Delta m_{31}^2 >0$ or inverted hierarchy (IH) when sign of $\Delta m_{31}^2 <0$. Likewise, if $\theta_{23}$ is not maximal, the question of the octant of $\theta_{23}$ refers to finding the correct octant of $\theta_{23}$ ($\theta_{23}<\pi/4$ is the lower octant (LO) and  $\theta_{23}>\pi/4$ is the higher octant (LO)).

Despite tremendous progress in our understanding of neutrino mass and mixing from several experiments, there are hints pointing towards physics BSM. If true, such hints could drastically impact our understanding/interpretation of neutrino mass and mixings. One such hint comes from a class of experiments referred to as the short baseline experiments. It was suggested that there could be oscillations driven by additional sterile neutrino state with mass-squared differences of order $1$ $\textrm{eV}^2$~\cite{Giunti:2019aiy,Dasgupta:2021ies}. The oscillations driven by $~1$ $\textrm{eV}^2$ mass splitting could leave their imprints at long baselines, thereby impacting the clean extrication of the unknowns in the neutrino oscillation physics. 

The first hint in favor of sterile neutrino came from the Liquid Scintillator Neutrino Detector (LSND), which reported an excess of $\bar{\nu}^{}_{e}$ events with a significance of $3.8\,\sigma$~\cite{LSND:1996ubh}. This observation was later supported by the MiniBooNE experiment, which reported the excess at $4.8\,\sigma$ significance~\cite{MiniBooNE:2020pnu}. The combined significance of the LSND and MiniBooNE brought the excess was around $6.1\, \sigma$ ~\cite{MiniBooNE:2020pnu}. Further, MicroBooNE experiment reported no evidence in support of the existence of the eV-scale sterile neutrino~\cite{MicroBooNE:2022sdp}.
A combined analysis of MiniBooNE and MicroBooNE data suggested a preference for oscillations involving three active  and one sterile neutrino state~\cite{MiniBooNE:2022emn}. It should also be noted that Gallium-based radiochemical experiments designed to study solar neutrinos, such as SAGE~\cite{SAGE:2009eeu}, GALLEX~\cite{Kaether:2010ag}, and the Baksan Experiment on Sterile Transition (BEST)~\cite{Barinov:2021asz} , observed a deficit in $\nu^{}_{e}$ events at a significance greater than $3\,\sigma$, providing hints of $\nu^{}_{e}\to\nu^{}_{s}$ oscillation with mass-squared difference, $\lldm \sim 1\,\text{eV}^{2}$. Similar support came from the reactor anti-neutrino anomaly (RAA)~\cite{Huber:2011wv, Mention:2011rk} and several experiments have been planned to test this anomaly. The Reactor-based Neutrino-4 experiment supported the active-sterile oscillation with $\lldm \sim 7\,\text{eV}^{2}$ and $\sin^{2}\td \sim 0.09$ at $3\,\sigma$ confidence level (C.L.)~\cite{Serebrov:2020kmd}. In future, the Short Baseline Neutrino (SBN) program~\cite{Machado:2019oxb} at Fermilab comprising of three detectors, Short-Baseline Near Detector (SBND)~\cite{SBND:2024vgn}, MicroBooNE~\cite{MicroBooNE:2022sdp} and Short-Baseline Far Detector (SBFD)~\cite{Torretta:2024fbn} is expected to clarify the situation regarding the existence of eV-scale sterile neutrino. 

{{There are several competing and complimentary long baseline experiments (with $L/E \sim {\cal O} (500)\, \textrm{km}/\textrm{GeV}$) that aim to improve the precision on measured parameters as well as to resolve some of the open questions in neutrino oscillation physics, for instance, Tokai to Hyper-Kamiokande (T2HK)~\cite{Hyper-Kamiokande:2018ofw} in Japan with a baseline of $295$ km and peak neutrino energy of $\sim 0.6$ GeV, Deep Underground Neutrino Experiment (DUNE) with a baseline of $1300$ km and peak neutrino energy of $\sim 2.5$ GeV~\cite{DUNE:2020ypp, DUNE:2021cuw} and Protvino to Orca (P2O)~\cite{Akindinov:2019flp, KM3NET:2016zxf} with a baseline of $2595$ km and peak neutrino energy of $\sim 5.1$ GeV. }}

A lot of effort has gone into studies pertaining to theoretical and phenomenological implications of an eV-scale sterile neutrino on signals at long baseline experiments~\cite{Goswami:1995yq,Gandhi:2015xza,Parke:2015goa,Dutta:2016glq,Gandhi:2017vzo,Kosmas:2017zbh,Choubey:2017cba,Agarwalla:2018nlx,Reyimuaji:2019wbn,Chatterjee:2022pqg, Singha:2022btw,Chattopadhyay:2022hkw,Majhi:2019hdj,Fiza:2021gvq,Parveen:2023ixk,Parveen:2024bcc}. 
In  earlier work involving some of the authors, we have addressed  the question of separating between the  $(3+0)$ and $(3+1)$ scenarios at different long baseline experiments such as T2HK, DUNE and P2O~\cite{Parveen:2023ixk}. We also showed that DUNE will set competetive constraints on the parameter space of active-sterile mixing parameters by using various experimental configurations such as high energy beam tunes - low energy (LE) and medium energy (ME), near detector (ND), far detector (FD), charged current (CC) interactions and neutral current (NC) interactions  etc~\cite{Parveen:2024bcc}. Most of these studies consider the energy resolution to be the standard as given in the technical reports of the respective experiments.

In the present study, we consider two physics scenarios : three active neutrinos i.e., the (3+0) case and three active plus one sterile neutrino i.e., the (3+1) case.  We take into account improved energy resolution~\cite{DeRomeri:2016qwo, Chatterjee:2021wac} over and above  the one detailed in DUNE Technical Design Report (TDR)~\cite{DUNE:2020ypp, DUNE:2021cuw} and demonstrate its impact on sensitivities to standard unknowns. 
The impact of energy resolution on a  particular new physics scenario of neutrino non-standard interactions (NSI) has been explored in~\cite{Chatterjee:2021wac}.
Our findings show that better neutrino energy resolution significantly enhances the DUNE sensitivity in both the  considered scenarios. The improved energy resolution leads to a better reconstruction particularly in the neighbourhood of first oscillation peak of DUNE.

The plan of the article is as follows. In Sec.~\ref{sec:framework}, we provide analytical discussion in  $(3+0)$ and $(3+1)$ case  with probability differences for CP violation and MH for the relevant appearance and disappearance channels. Sec~\ref{sec:setup} provides details of the experiment and simulation with details about energy resolution. In Sec.~\ref{sec:event}, we give event rates obtained using DUNE TDR and best reconstruction scenario. Sec.~\ref{sec:analysis} provides details of the $\chi^2$ sensitivities for the three unknowns : 
 CP violation, MH and octant of $\theta_{23}$. The results and discussions are presented in Sec.~\ref{sec:result}.  We summarize in Sec.~\ref{sec:conclusion}.

\section{Theoretical framework} 
\label{sec:framework}
\subsection{Hamiltonian in the $(3+0)$ case} 
\label{sec:ham_std}

In the three flavor framework, the neutrino mixing matrix in the commonly adopted Pontecorvo–Maki–Nakagawa–Sakata (PMNS) form is~\cite{Pontecorvo:1957qd,Pontecorvo:1957cp, Maki:1962mu,Gribov:1968kq,ParticleDataGroup:2024cfk}
\small{\bea 
\mathcal{U} = 
\begin{pmatrix}
c^{}_{12}c^{}_{13} & s^{}_{12}c^{}_{13} & s^{}_{13} \text{e}^{-i\delta_{13}}\\
-s^{}_{12}c^{}_{23}-c^{}_{12}s^{}_{13}s^{}_{23} \text{e}^{i\delta_{13}} & c^{}_{12}c^{}_{23}-s^{}_{12}s^{}_{13}s^{}_{23} \text{e}^{i\delta_{13}} & c^{}_{13}s^{}_{23} \\
s^{}_{12}s^{}_{23}-c^{}_{12}s^{}_{13}c^{}_{23} \text{e}^{i\delta_{13}} & -c^{}_{12}s^{}_{23}-s^{}_{12}s^{}_{13}c^{}_{23} \text{e}^{i\delta_{13}} & c^{}_{13}c^{}_{23}
\end{pmatrix} 
\begin{pmatrix}
1 & & \\
& \text{e}^{i\phi_{1}}&
 \\
& & \text{e}^{i\phi_{2}}
\end{pmatrix} \,,
\label{eq:PMNS}
\eea }\noindent
where, $c^{}_{ij}=\cos\theta^{}_{ij}$ and $s^{}_{ij}=\sin\theta^{}_{ij}$. Note that only the Dirac phase ($\delta_{13}$) affects neutrino oscillations and  the two Majorana phases ($\phi_1$ and $\phi_2$) have no effect on neutrino oscillations.

We can write down the effective Hamiltonian in flavor basis as
\bea
\mathcal{H} &=& \mathcal{U}\left[\dfrac{1}{2E}\begin{pmatrix}0 & 0 & 0\\0 & \alpha & 0\\0 & 0 & 1 \end{pmatrix}\right] \mathcal{U}^\dagger + \begin{pmatrix} V^{}_{{CC}} & 0 & 0\\0 & 0 & 0\\0 & 0 & 0\end{pmatrix}\,,
\eea
where, $\Delta m^{2}_{21}= m^{2}_{2}-m^{2}_{1}$, $\Delta m^{2}_{31} =m^{2}_{3}-m^{2}_{1}$,  and 
$\alpha = \Delta m^{2}_{21}/{\Delta m^{2}_{31}}$. $V^{}_{\textrm{CC}}$ is the matter-induced potential due to the standard  CC interactions of $\nu^{}_{e}$ with electrons, given by
\bea
V^{}_{{CC}}=\sqrt{2}\,G_{F} N_{{e}} = 7.6 \times 10^{-14}\,Y_{{e}} \,\left[\dfrac{\rho}{\rm{g/cc}}\right]\,\textrm{eV}\,, \nonumber
\eea \noindent
where $G_F$ is the Fermi coupling constant  and $N_{e} = N_{\rm{Avo}}\,\rho\, Y_{{e}}$ is the electron number density, $\rho$ is the mass density,  $N_{\rm{Avo}} = 6.023 \times 10^{23} \rm{g}^{-1} \rm{mole}^{-1}$ is the Avogadro's number and  $Y_{e} \simeq 0.5$ is  the number of electrons per nucleon.

\subsection{Hamiltonian for the $(3+1)$ case} 
\label{sec:ham_str}

 In the $(3+1)$ framework, the mixing matrix $\mathcal{U}$ is a $4 \times 4$ unitary matrix (different parameterizations have been used in literature such as~\cite{Kopp:2013vaa, Klop:2014ima, DUNE:2020fgq}). In the present work, we adopt the notation~\cite{Klop:2014ima}.
\bea
\mathcal{U} &=&V(\theta_{34}, \delta_{34})O(\theta_{24})V(\theta_{14}, \delta_{14})O(\theta_{23})V(\theta_{13}, \delta_{13})O(\theta_{12})\,, \nonumber 
\label{eq:umat}
\eea \noindent
where, $V(\theta_{ij},\delta_{ij})$ and $O(\theta_{ij})$ are complex and real rotation matrices  in the $i$-$j$ plane respectively (with $i , j = 1, 2, 3, 4~\textrm{and}~i \neq j$). 

The effective Hamiltonian for the $(3+1)$ case in flavor basis is
\bea
\mathcal{H} &=& \mathcal{U} \left[\dfrac{1}{2E}\begin{pmatrix}0 & 0 & 0&0\\0 & \alpha & 0&0\\0 & 0 & 1&0\\0 & 0 & 0&R \end{pmatrix}\right] \mathcal{U}^\dagger + \begin{pmatrix}V^{}_{{CC}}+V^{}_{{NC}} & 0 & 0&0\\0 & V^{}_{{NC}}  & 0&0\\0 & 0 & V^{}_{{NC}} &0\\0 & 0 & 0&0\end{pmatrix}\, ,
\eea \noindent
where $R = {\Delta m^{2}_{41}}/{\Delta m^{2}_{31}}$ is a dimensionless quantity and $V_{\textrm{NC}}=-\sqrt{2} G_F N_{n}/{2}$ is the NC potential. Note that in the $(3+1)$ framework, there are three new angles ($\theta_{14}, \theta_{24}, \theta_{34}$), two new phases ($\delta_{14}, \delta_{34}$) and one additional squared-mass difference $\lldm$.

In what follows, we present observables to quantify CP violation and MH in the two scenarios.

\subsection{CP violating probability differences for the $(3+0)$ and $(3+1)$ cases}
\label{subsec:CP}
The CP violating probability difference is given by
 \bea
     \Delta {P}_{\alpha\beta}^{CP}&=& {{P}_{\alpha \beta} - \bar {P}_{\alpha \beta}}\,, 
     \label{eq:CP}
 \eea \noindent
where $P_{\alpha \beta}$ is the probability for $\nu_{\alpha}\to\nu_{\beta}$ and $\bar{P}_{\alpha \beta}$ is the 
 probability for antineutrinos.

We use some  approximations for calculating the $\Delta P^{CP}_{\mu e}$ in the $(3+1)$ case as mentioned in~\cite{Klop:2014ima}. We average over oscillations induced by the largest mass-squared difference,  $\lldm$. Since $\theta_{13} , \theta_{14} , \theta_{24} \leq 13^\circ$~\cite{Dentler:2018sju},  we have set $\sin\theta_{ij} \sim 10^{-1} \sim \lambda$, where $\theta_{ij}$ could be $\theta_{13}$, $\theta_{14}$ or $\theta_{24}$. 

For the appearance channel $(\nu_{\mu}\to\nu_{e})$, we have 
\begin{subequations}
 \bea
  \Delta P^{CP(3+1)}_{\mu e}(\td,\te)  &\simeq& \Delta P^{CP(3+0)}_{\mu e}+ 4 s_{13} s_{14} s_{24} s_{23}  \Bigg\{\sin[\da-\db]\Bigg[\dfrac{\sin[\hat{A}-1]\Delta}{(\hat{A}-1)}\cos[(1-\hat{A})\Delta]\nonumber \\ &+&\dfrac{\sin[\hat{A}+1]\Delta}{(\hat{A}+1)}\cos[(1+\hat{A}) \Delta]\Bigg]+\cos[\da-\db]\Bigg[\dfrac{\sin[\hat{A}-1]\Delta}{(\hat{A}-1)} \sin[(1-\hat{A}) \Delta]\nonumber \\ &-&\dfrac{\sin[\hat{A}+1]\Delta}{(\hat{A}+1)}\sin[(1+\hat{A}) \Delta]\Bigg]\Bigg\} + \mathcal{O}(\lambda^{4})\,,      
\label{eq:Delta P_mue_4nu}
  \eea \noindent
Note that the expression for $P^{{CP}(3+0)}_{\mu e}$ is given in Appendix~\ref{app_1}.
For the disappearance channel $(\nu_{\mu}\to\nu_{\mu})$, we have
   \bea
\Delta P^{{CP(3+1)}}_{\mu \mu} (\td,\te) &\simeq& \Delta P^{{CP}(3+0)}_{\mu \mu}
 + \frac{1}{4}s_{13}s_{14}s_{24}c_{24}\Bigg\{-s_{23}\bigg[\frac{\cos[\delta_{14}+(\hat{A}-1)\Delta]}{\hat{A}-1} \nonumber \\
 &+& \frac{\cos[\delta_{14}+(\hat{A}+1)\Delta]}{\hat{A}+1}\bigg]
 +\frac{2\hat{A}}{\hat{A}-1}s_{23}\cos[\delta_{14}+2\hat{A}\Delta]\, \nonumber \\
 &+& 2\sin[\delta_{13}+\delta_{14}]\sin2\theta_{23}\left[\frac{\sin[2\Delta+\delta_{13}]}{\hat{A}+1}-\frac{\sin[2\Delta-\delta_{13}]}{\hat{A}-1}\right] + 2\sin3\theta_{23}\nonumber \\&&\left[\frac{\cos[2\Delta-\delta_{13}]\cos\big[\delta_{14}-\delta_{13}/2-\omega_{-}\big]}{\hat{A}-1}+\frac{\cos[2\Delta+\delta_{13}]\cos\big[\delta_{14}-\delta_{13}/2+\omega_{+} \big]}{\hat{A}+1}\right] \nonumber \\
&+& 2\sin 4\theta_{23}\cos[\delta_{14}+\delta_{13}]\Big[\cos(2\Delta-\delta_{13}]\Omega_{-} + \cos[2\Delta+\delta_{13}]\Omega_{+} \Big]
 \Bigg\} \nonumber \\
&+& \frac{1}{2} s_{14} s_{13} s_{23} s_{24} c_{24}\Bigg\{\frac{\cos[\delta_{14}-2\Delta]}{\hat{A}-1}-\frac{\cos[\delta_{14}+2\Delta]}{\hat{A}+1}-\cos[\delta_{14}-\delta_{13}]\bigg(\frac{2\hat{A}}{\hat{A}^2-1}\bigg) \nonumber \\
&+& 2\cos2\theta_{23}\Big[\Omega_{-} \cos[\delta_{14}-\Delta-\delta_{13}/2]+ \Omega_{+} \cos[\delta_{14}+\Delta-\delta_{13}/2]\Big]\Bigg\} + \mathcal{O}(\lambda^{4}) \,, 
\label{eq:Delta P_mumu_4nu}
  \eea \noindent
\end{subequations}
where, we have used
\bea
&&
 \Delta =  \dfrac{\Delta m^{2}_{31} L} {4E},~~\hat {A} = \dfrac{A}{\Delta m^{2}_{31}},~~\Omega_{\pm} =  \frac{\cos(\Delta \pm \delta_{13}/2)}{\hat A \pm 1},~~\omega_{\pm} = (1 \pm 2 \hat A ) \Delta\,. \nonumber
\eea \noindent
Note that the expression for $P^{{CP}(3+0)}_{\mu \mu}$ is given in Appendix~\ref{app_1}.
   \begin{figure}[t!]
\centering
\includegraphics[width=7in]{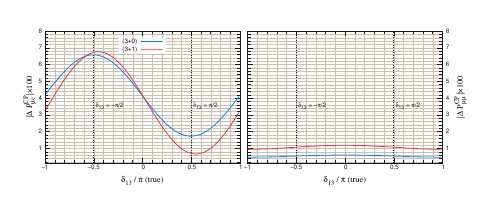}
\caption{\footnotesize{$|\Delta P^{CP}_{\mu e}|$ and $|\Delta P^{CP}_{\mu \mu}|$   plotted as a function of $\delta_{13}$ for $L = 1300$ km, $E = 2.6$ GeV. The  $(3+0)$  and $(3+1)$ cases are depicted by blue and red curves respectively.}}
\label{fig:cp}
\end{figure} \noindent
 \begin{table}[h]
 \arrayrulecolor{black}
\centering
\scalebox{0.8}{
\begin{tabular}{| c | c | c | c |}
\hline
\rowcolor{blue!5}Parameter & Best-fit value & 3$\sigma$ interval & $1\sigma$ uncertainty    \\ [2mm]
\hline 
$\theta^{}_{12}/^\circ$            & $34.3$                    &  $31.4$ - $37.4$  &  $2.9\%$ \\[2mm]
\rowcolor{blue!5} $\theta^{}_{13}/^\circ$    & $8.53$ $(8.58)$              &  $8.13$  -  $8.92$  ($8.17$  -  $8.96$) &  $1.5\%$ \\ [2mm]
 $\delta^{}_{13}/^\circ$   & $194$ ($284$)    & [$128$ - $359$] \big([$200$ - $353$]\big) & - \\[2mm]
 \rowcolor{blue!5}$\ldm$ [$\text{eV}^2$] & $+2.55$ ($-2.45$) $\times 10^{-3}$   &  [$2.47$ - $2.63$] $\big(-[2.37 - 2.53]\big)\times 10^{-3}$ & $1.2\%$  \\[2mm]
$\sdm$ [$\text{eV}^2$]  & $7.5 \times 10^{-5}$  &  [$6.94$ - $8.14$] $\times 10^{-5}$ &  $2.7\%$  \\[2mm]
\rowcolor{blue!5}$\lldm$ [$\text{eV}^2$]  & $1$ & - & \\ [2mm]
$\td /^\circ$  & $5.7$ & $0$ - $18.4$ & $\sigma(\sin^{2}\theta_{14}) = 5\%$ \\[2mm]
\rowcolor{blue!5}$\te /^\circ$  & $5$ & $0$ - $6.05$ & $\sigma(\sin^{2}\theta_{24}) = 5\%$ \\[2mm]
$\tf /^\circ$  & $20$ & $0$ - $25.8$ & $\sigma(\sin^{2}\theta_{34}) = 5\%$ \\[2mm]
\rowcolor{blue!5}$\delta_{14}/^\circ$  & $0$   & [$-180$, $180$] &  - \\[2mm]
$\delta_{34} /^\circ$  & $0$   & [$-180$, $180$] &  - \\[2mm]
\hline
\end{tabular}}
\vskip 0.1in
\caption{\footnotesize{\label{tab:parameters} 
The values of the oscillation parameters and their uncertainties. The values of standard (3+0) parameters have been taken from the global fit analysis~\cite{deSalas:2020pgw} while the sterile (3+1) parameter values have been chosen from~\cite{Dentler:2018sju}.  
If the $3\sigma$ upper and lower limit of a parameter is $x_{u}$ and $x_{l}$ respectively, the $1\sigma$  uncertainty is $(x_{u}-x_{l})/3(x_{u}+x_{l})\%$~\cite{DUNE:2020ypp}. 
For the active-sterile mixing angles, a conservative $5\%$ uncertainty has been used on $\sin^{2}\theta_{i4}$ ($i = 1, 2, 3$). The bracketed values correspond to the case when the hierarchy is inverted. All other parameter values are for NH.
}}
\end{table}   
\noindent
We retain terms upto order $\mathcal{O}(\lambda^{3})$. In the limiting case, when $(\theta_{14},\theta_{24},\theta_{34},\delta_{14},\delta_{34})$ tend to zero, we  recover the corresponding probability differences in $(3+0)$ case~\cite{Akhmedov:2004ny, Parveen:2023ixk}.

 In Fig~\ref{fig:cp}, we plot the CP violating differences, i.e., $\Delta P^{{CP}}_{\alpha \beta}$, as a function of $\delta_{13}$  for $L = 1300$ km, $E = 2.6$ GeV for the two scenarios under consideration. 
 The left panel  is for $\nu_{\mu}\to\nu_{e}$ channel where as the right panel corresponds to  $\nu_{\mu}\to\nu_{\mu}$ channel. The simulations have been performed using the software General Long Baseline Experiment Simulator (GLoBES)~\cite{Huber:2004ka,Huber:2007ji} which solves the full three flavor neutrino propagation equations numerically using the Preliminary Reference Earth Model (PREM)~\cite{Dziewonski:1981xy} density profile of the Earth. 
 
\subsection{MH probability differences for the $(3+0)$ and $(3+1)$ cases}
 \label{subsec:mh}
 The MH probability difference is given by
 \bea
     \Delta {P}_{\alpha\beta}^{{MH}}&=& {{P}^{{NH}}_{\alpha \beta} - {P}^{{IH}}_{\alpha \beta}}\,, 
     \label{eq:MH}
 \eea \noindent
where $P^{{NH}}_{\alpha \beta}$ is the probability for NH and $P^{{IH}}_{\alpha \beta}$ is the probability in case of IH. 
Analytic expressions for probability for NH are given in~\cite{Klop:2014ima}. By replacing $\Delta$ with $- \Delta$, $\alpha$ with $- \alpha$ and $\hat{A}$ with $-\hat{A}$ in the probability expressions for NH, we can obtain the expressions for IH. Thus, MH probability difference for the  appearance channel $(\nu_{\mu}\to\nu_{e})$ is given by
\begin{subequations}
\bea
\Delta P^{{MH(3+1)}}_{\mu e}(\theta_{14},\theta_{24}) &\simeq&   \Delta P^{{MH}(3+0)}_{\mu e}+ 4 s_{13} s_{14} s_{24} s_{23}  \Bigg\{\sin[\da-\db]\Bigg[\dfrac{\sin[\hat{A}-1]\Delta}{(\hat{A}-1)} \cos[(1-\hat{A}) \Delta]\nonumber \\ &+&\dfrac{\sin[\hat{A}+1]\Delta}{(\hat{A}+1)}\cos[(1+\hat{A}) \Delta]\Bigg]+\cos[\da-\db]\Bigg[\dfrac{\sin[\hat{A}-1]\Delta}{(\hat{A}-1)} \sin[(1-\hat{A}) \Delta]\nonumber \\&-&\dfrac{\sin[\hat{A}+1]\Delta}{(\hat{A}+1)}\sin[(1+\hat{A}) \Delta]\Bigg]\Bigg\} +  \mathcal{O}(\lambda^{4})\,, 
 \label{eq:Delta P_mue_4nu_mh} 
\eea \noindent
It may be noted that the expression for $P^{{MH}(3+0)}_{\mu e}$ is given in Appendix~\ref{app_1}.
For the disappearance channel $(\nu_{\mu}\to\nu_{\mu})$, we have 
\bea
\Delta P^{{MH(3+1)}}_{\mu \mu}(\theta_{14},\theta_{24}) &\simeq& \Delta P^{{MH(3+0)}}_{\mu \mu}
 + \frac{1}{4} s_{13} s_{14} s_{24}c_{24}\Bigg\{-3 s_{23}\Bigg[ \frac{\cos[ \delta_{14} + (\hat {A}-1) \Delta]}{\hat A-1}\nonumber \\ &+&
 \frac{\cos [\delta_{14} - (\hat {A}+1) \Delta]}{\hat A+1} \Bigg] + \dfrac{2 \hat{A}}{(\hat{A}^{2}-1)}  s_{23} \cos[\delta_{14} + 2 \hat {A}\Delta + \delta_{13}]-2\sin[\delta_{14} + \delta_{13}]\nonumber \\&&\sin 2 \theta_{23}\left[\frac{\sin[2\Delta - \delta_{13}]}{\hat{A}-1} -\frac{\sin[2\Delta +\delta_{13}]}{\hat{A}+1}\right]+ 2 \sin 3 \theta_{23}\nonumber \\&& \left[\frac{\cos[\delta_{14} - \omega_{-} - \delta_{13} / 2]\cos [2 \Delta - \delta_{13}]} {\hat {A}- 1} + \frac{\cos[\delta_{14} - \omega_{+} -\delta_{13} / 2]\cos [2 \Delta +\delta_{13}]} {\hat {A}+1} \right] \Bigg\}\nonumber \\ &+& 2 \cos [\delta_{14} + \delta_{13}] \sin 4 \theta_{23} \Big[\Omega_{-}\cos [2\Delta-\delta_{13}] +\cos [2\Delta+\delta_{13}] \Omega_{+} \Big] + \frac{1}{2} s_{14} s_{13} s_{23} s_{24} c_{24} \nonumber \\ && \Bigg\{\left[\frac{\cos [\delta_{14} - 2 \Delta]}{\hat A -1} + \frac{\cos [\delta_{14} + 2 \Delta]}{\hat A +1}\right]- 2 \cos[\delta_{14}- \delta_{13}]\Big[\Omega_{-} \cos[\delta_{14}-\Delta-\delta_{13}/2]\nonumber \\ &+& \Omega_{+} \cos[\delta_{14}+\Delta-\delta_{13}/2]\cos 2 \theta_{23}\Big]\Bigg\} +\mathcal{O}(\lambda^{4})\,.
\label{eq:Delta P_mumu_4nu_mh} 
  \eea \noindent
  \end{subequations}
\begin{figure}[t!]
\centering
\includegraphics[width=7in]{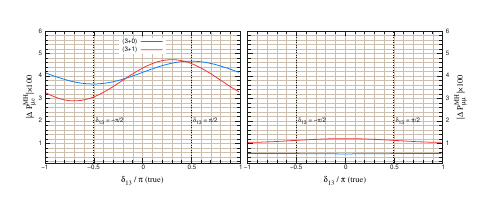}
\caption{\footnotesize{$|\Delta P^{MH}_{\mu e}|$ and $|\Delta P^{MH}_{\mu \mu}|$   plotted as a function of $\delta_{13}$ for $L = 1300$ km, $E = 2.6$ GeV. The  $(3+0)$  and $(3+1)$ cases are depicted by blue and red curves respectively.}}
\label{fig:mh}
\end{figure} \noindent
 
The expression for $P^{{MH}(3+0)}_{\mu \mu}$ is given in Appendix~\ref{app_1}. 
In Fig~\ref{fig:mh}, we plot the MH probability differences, i.e., $\Delta P^{{MH}}_{\alpha \beta}$, as a function of $\delta_{13}$  for $L = 1300$ km, $E = 2.6$ GeV for the two scenarios under consideration.

\section{Details of the Experiment and Simulation} 
\label{sec:setup}

\subsection{Experiment details}

We use the latest configuration of the DUNE TDR~\cite{DUNE:2020ypp,DUNE:2021cuw} in the present study. Neutrino beam originates at Fermi National Accelerator Laboratory (Fermilab) in Batavia, Illinois. The ND of DUNE, with a target mass of $0.067$ kt, will be installed $570$ m downstream. The FD, with a target mass of approximately 40 kt, will be located $1.5$ km underground at the Sanford Underground Research Facility (SURF) in South Dakota, at a baseline of around $1300$ km from the neutrino source. The DUNE experiment will utilize a $120$ GeV proton beam of  $1.2$ MW power with total runtime of $13$ years equally divided into neutrino and anti-neutrino mode. The total exposure is taken to be $624$ kt$\cdot$MW$\cdot$years.

\begin{figure}[t!]
   \centering
\includegraphics[width=4.2in]{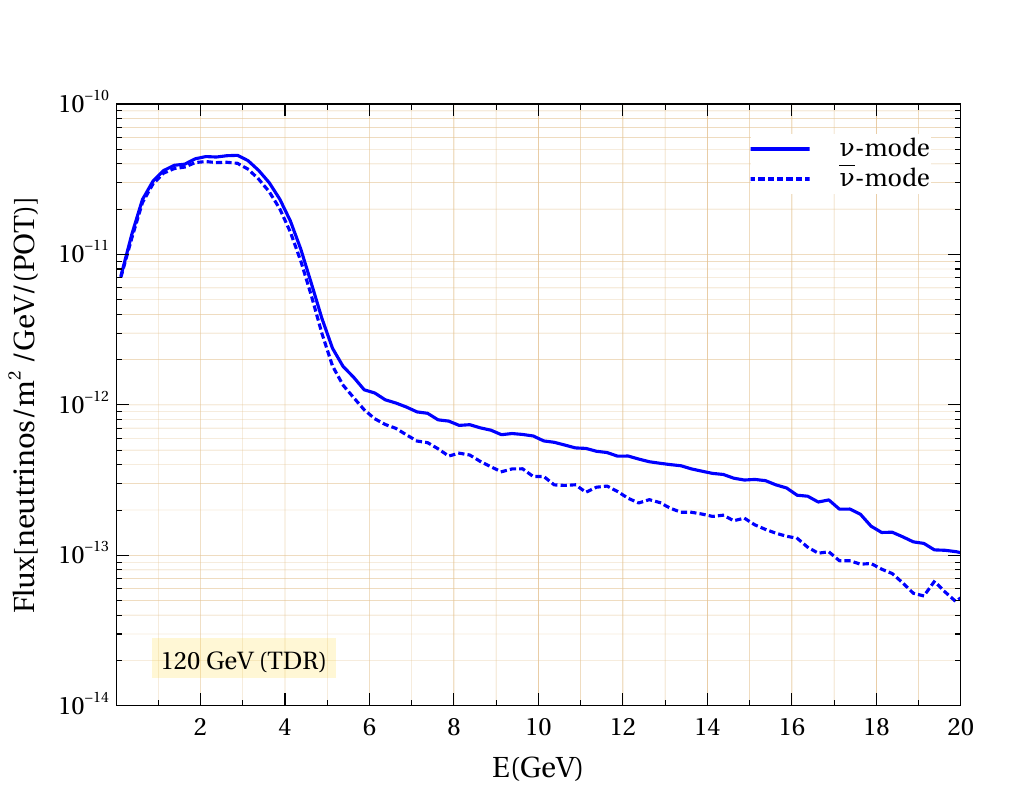}
\caption {\footnotesize{The blue curve represents the flux corresponding to DUNE TDR LE beam~\cite{DUNE:2020ypp,DUNE:2021cuw}. The solid and
dotted curves indicate the $\nu_{\mu}$ and $\bar \nu_{\mu}$ flux respectively.}}
\label{fig:dune_flux}
\end{figure}

We use the standard LE beam which peaks around $\sim 2$ - $3$ GeV~Fig.~\ref{fig:dune_flux}.
The neutrino beam is generated by directing a 120 GeV proton beam onto a graphite target. The resulting hadronic interactions are simulated using G4LBNF, a GEANT4-based framework~\cite{Agostinelli:2002hh,Allison:2006ve} developed for modeling the long baseline neutrino facility (LBNF) beamline~\cite{DUNE:2020ypp}. The secondary hadrons produced in the target are focused by a system of three magnetic horns, each operating at a current of 300 kA. These hadrons are then allowed to decay within a 194 meter long, helium-filled decay pipe, yielding the LE neutrino flux. The focusing horns can be operated in forward and reverse current configurations to produce $\nu$ (solid blue) and $\bar{\nu}$ (dotted blue) beams respectively (see Fig.~\ref{fig:dune_flux}). 

\subsection{Simulation details} 

The simulations and sensitivity estimates have been carried out using the GLoBES software package~\cite{Huber:2004ka,Huber:2007ji}. We numerically solve the full three flavor neutrino propagation equations assuming the Earth's matter density profile given by the Preliminary Reference Earth Model (PREM)~\cite{Dziewonski:1981xy}.\footnote{Although it is possible to incorporate uncertainties in the Earth's matter density, previous studies~\cite{Gandhi:2004md, Gandhi:2004bj, Kelly:2018kmb, Chatterjee:2018dyd} have shown that such variations have  insignificant impact on the results.} For true values of oscillation parameters, their $3\,\sigma$ ranges and corresponding uncertainties, we use the values given in Table~\ref{tab:parameters}. We assume a $2\% $  uncertainty in the matter density~\cite{DUNE:2021cuw} and marginalize over it when computing the minimum $\chi^2$.

     \noindent

\subsection{Improved energy resolution}
\label{sec:resolution}
DUNE is based on liquid argon time projection chamber (LArTPC) technology~\cite{Abi:2020loh}. The neutrino interacts with an argon nucleus in the neutrino detector and produces charged as well as neutral particles. The charged particles
travel through the liquid argon and loose their energy through ionization and excitation of argon atoms. The ionized electrons propagate towards a set of anode wire planes under the influence of electric field and are finally collected. This charge deposition helps in obtaining showers and tracks which in turn leads to particle identification. The particle identification and energy deposition through the ionization together allows us to reconstruct the energy of individual particle, thereby allowing to infer the energy of the incoming neutrino. 

\begin{figure}[ht!]
   \centering
\includegraphics[width=6.5in]{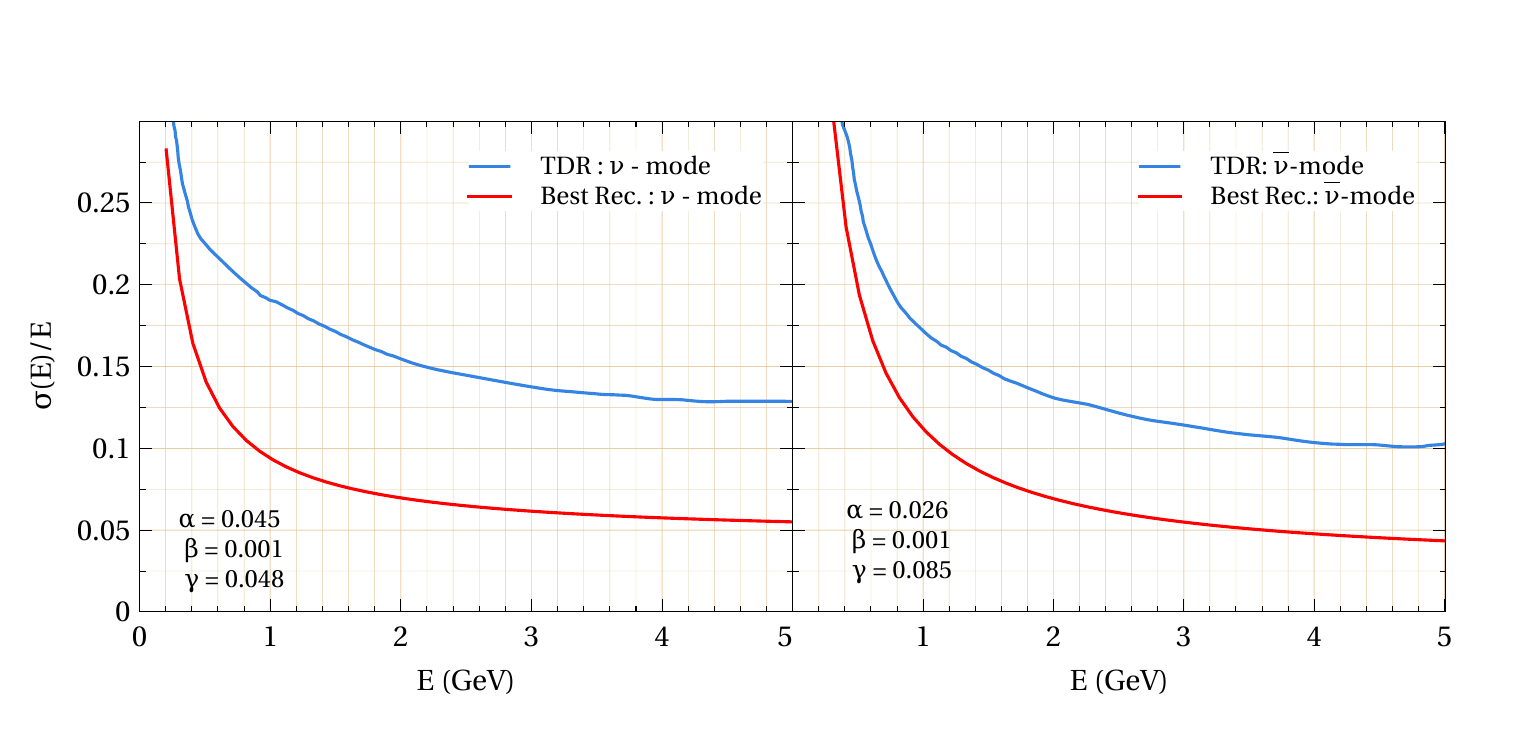}
\caption {\footnotesize{Energy resolution plotted as a function of neutrino energy. The blue and red curves depict the energy resolutions for DUNE TDR~\cite{DUNE:2020ypp} and best reconstruction scenario respectively~\cite{Chatterjee:2021wac}.  The left (right) panel corresponds to $\nu$ ($\bar \nu$) - mode. }}
\label{fig:res}
\end{figure}

In this work, we consider the parametrization used in~\cite{Chatterjee:2021wac} for the energy resolution function
\bea
R(E,E_r)=e^{-(E-E_r)^2/2\sigma^2}/\sigma\sqrt{2\pi} \, ,
\eea 
where $E$ is the true neutrino energy, $E_r$ is the reconstructed energy, and the energy resolution $\sigma$ is given by
\bea
\sigma(E)/{\rm GeV} &=& \alpha \cdot (E/{\rm GeV}) + \beta \cdot \sqrt{E/ {\rm GeV}} + \gamma \, ,
\label{eq:ene_res}
\eea
with fit parameters $(\alpha,\,\beta,\,\gamma)$ are $(0.045,\, 0.001,\, 0.048)$ for neutrinos and $(0.026,\,0.001,\,0.085)$ for antineutrinos (Fig.~\ref{fig:res}). This above values of  fit parameters $(\alpha,\,\beta,\,\gamma)$ are obtained from the best reconstruction case given in ~\cite{Friedland:2018vry}.

{In~\cite{Friedland:2018vry}, a detailed simulation has been performed by identifying different channels where energy is lost or missed, such as sub-threshold particles, neutral particles like neutrons that deposit energy inefficiently, energy consumed in nuclear processes, and charge recombination effects. Such energy losses cause fluctuations in the visible energy, which limit the achievable energy resolution. Therefore, improving particle identification, lowering detection thresholds, and accounting for charge recombination improves the energy resolution. The Fig.~\ref{fig:res} presents the energy resolution for neutrinos (left panel) and anti-neutrinos (right panel) as a function of energy for the DUNE TDR case (blue curve) and the best reconstruction scenario (red curve).}

In the present work, our focus is to consider improved energy resolution and study its impact on the sensitivities to current unknowns for the $(3+0)$ case and  $(3+1)$ case at DUNE. For all the simulation studies pertaining to the best reconstruction case, we take the  resolution for both appearance and disappearance channels to be the same. The  migration matrices used for $NC$, $\nu_e$ contamination, misidentified muon, and $\nu_\mu\to\nu_\tau$ backgrounds are taken from~\cite{DUNE:2021cuw}.
  \noindent
\section{Event rates}
\label{sec:event}
\begin{figure}[t!]
\centering
\includegraphics[width=7in]{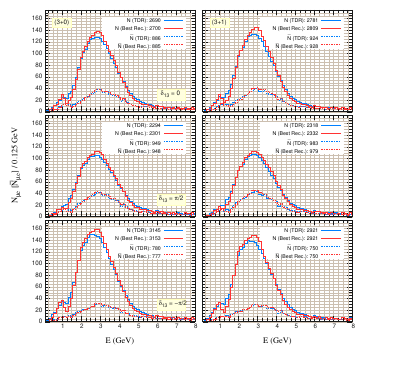}
\caption{\footnotesize{$N_{\mu e}\,(\bar N_{\mu e})$ plotted as a function of energy for DUNE TDR and best reconstruction scenario for the $(3+0)$ and $(3+1)$ scenario at DUNE. The neutrino MH is assumed be NH.}}
\label{fig:mue_events}
\end{figure} \noindent
\begin{figure}[t!]
\centering
\includegraphics[width=7in]{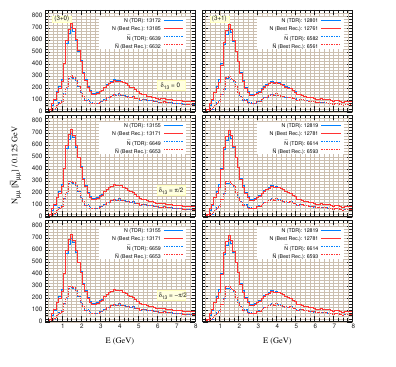}
\caption{\footnotesize{Same as Fig.~\ref{fig:mue_events} but for $\nu_{\mu}\to\nu_{\mu}$ $(\bar {\nu}_{\mu}\to \bar{\nu}_{\mu})$ channel.}}
\label{fig:mumu_events} 
\end{figure} \noindent

Theoretically, the expected event rate is given by
\bea
N_{\alpha \beta}~(L) &=& N_{target} \times  \int  
 \Phi_{\nu_\alpha} (E,L)  \times P_{\alpha \beta} (E,L) \times \sigma_{\nu_\beta} (E)~dE \, ,
\label{eq:eventeqn}
\eea
 where
  $N_{target}$ is the number of target nucleons per kt of detector fiducial volume, for DUNE, $N_{target}$ = $6.022 \times 10^{32}~N/\textrm{kt}$ where, $N$ stands for the number of nucleons.
  $P_{\alpha \beta}  (E,L)$ is the oscillation probability in matter, 
    $\Phi_{\nu_\alpha} (E,L)$ is the flux of $\nu_\alpha$, $\sigma _{\nu_\beta} (E)$ is the CC cross section of $\nu_\beta$. 
$\sigma _{\nu_e}(E)$ is given by~\cite{Bass:2013vcg, Masud:2016nuj} 
\bea
\label{eq:cross_sec}
 \sigma_{\nu_e}(E) &=& 0.67 \times 10^{-42} (m^2/{\rm{GeV}}/N) \times E\,, \quad {\rm {for}} \quad E > 0.5\,{\rm{GeV}}.
\eea
It may be noted that $\sigma_{\nu_{\mu}} \sim \sigma_{\nu_{e}}$, for the energy range, $0$ - $8$ GeV~\cite{Messier:1999kj}. For anti-neutrinos, the cross-section is roughly a factor of two smaller than the neutrinos~\cite{DUNE:2021cuw}. To compute the events, we use the  GLoBES software~\cite{Huber:2004ka,Huber:2007ji}. The neutrino fluxes have been taken from DUNE TDR~\cite{DUNE:2021cuw} (see Fig.~\ref{fig:dune_flux}). 

We generate event rate as a function of energy for $\nu_\mu \to \nu_e\,(\bar \nu_\mu \to \bar \nu_e)$ and  $\nu_\mu \to \nu_\mu \, (\bar \nu_\mu \to \bar \nu_\mu)$. The backgrounds include: (a) $\nu_{\mu}\to\nu_{e}/\nu_{\tau}$ and $\nu_{\mu}\to\nu_{\mu}$ CC in $\nu$ and $\bar{\nu}$-mode and (b) $\nu_\mu / \nu_e \to X$ (NC) for appearance and disappearance channels (see details in~\cite{DUNE:2021cuw}).

In Fig.~\ref{fig:mue_events} and Fig.~\ref{fig:mumu_events}, 
we present the event rates as a function of energy for TDR  and best reconstruction scenario for $\nu_\mu \to \nu_e\,(\bar \nu_\mu \to \bar \nu_e)$ and $\nu_\mu \to \nu_\mu \, (\bar \nu_\mu \to \bar \nu_\mu)$ respectively. We consider three different values of CP phase, $\delta_{13}= 0$ (first row), $\delta_{13}=\pi/2$ (second row) and $\delta_{13}=-\pi/2$ (third row). 
We note the following:
\begin{itemize}
\item $\nu_{\mu}\to\nu_{e}$ $(\bar {\nu}_{\mu}\to \bar{\nu}_{e})$ channel:
In Fig.~\ref{fig:mue_events}, the event spectrum for the two cases (TDR and best reconstruction) has been plotted as a function of energy for $(3+0)$ and $(3+1)$ cases taking fixed values of $\delta_{13} = 0, \pm \pi/2$. It can be noted that better reconstruction leads to larger number of events for all values of $\delta_{13}$. Larger deviation is seen in case of neutrinos for NH.

\item $\nu_{\mu}\to\nu_{\mu}$ $(\bar {\nu}_{\mu}\to \bar{\nu}_{\mu})$ channel: 
In Fig.~\ref{fig:mumu_events}, the event spectrum for the two cases (TDR and best reconstruction) have been plotted as a function of energy for $(3+0)$ and $(3+1)$ cases taking fixed values of $\delta_{13} = 0, \pm \pi/2$.
{For $\delta_{13} = 0$, there is a difference between the $(3+0)$ and $(3+1)$ cases ($\sim 371$ for TDR and $\sim 424$ for best reconstruction) in $\nu$-mode and ($\sim 57$ for TDR and $\sim 71$ for best reconstruction) in $\bar\nu$-mode. As expected, for $\delta_{13} = \pm \pi/2$, the event rates are largely the same for TDR and best reconstruction case respectively. }

\end{itemize}
\section{Analysis procedure}
  \label{sec:analysis}

To evaluate the sensitivities of DUNE to standard unknowns (CP violation, MH, and the octant of $\theta_{23}$), we perform the standard $\chisq$  analysis. Although the results are primarily obtained through numerical computations 
using the GLoBES software, analytical expressions of $\chisq$ for each of these quantities allow us to understand the sensitivity estimates.

{\bf{Sensitivity to CP violation:}}

Here we address the question of sensitivity of a given experiment to discriminate between CP conserving ($=0,\pi$) and CP violating ($\neq 0, \pi$) values of the Dirac CP phase $\delta_{13}$. 
Including only statistical effects, the $\chisq$ corresponding to CP violation sensitivity for a given oscillation channel is given by~\cite{Masud:2015xva, Rout:2020emr, Rout:2020cxi}
\bea
\chi^2 &\equiv&  \min_{\delta_{test}}  \sum_{i=1}^{x}  \sum_{j=\nu, \bar\nu}^{} 
 \frac{\left[N_{true}^{i,j}(\delta_{true}) - N_{test}^{i,j} (\delta_{test}=0,\pi )\right]^2 }{N_{true}^{i,j} (\delta_{true})}\,,
 \label{eq:chisq_cp}
\eea \noindent
where $N_{true}^{i,j}$ and $N_{test}^{i,j}$ represent the true and test datasets in the $\{i,j\}$-th bin respectively.  
Since no statistical fluctuations are introduced in the simulated data, we define $\Delta \chi^2 = \chi^2$, following Pearson’s definition of $\chisq$~\cite{Qian:2012zn}. 
The label $i$ stands for the energy bins, where $i = 1$ to $x$, and $x$ is the total number of bins, specific to the experiment under consideration. For DUNE experiment, there are $x=62$ energy bins each having a width of $0.125$ GeV in the energy range of $0.5$ - $8$ GeV and $2$ bins of width $1$ GeV each in the range $8$ - $10$ GeV~\cite{DUNE:2021cuw}. The label $j$ is being summed over the $\nu$ and $\bar \nu$-mode. To evaluate the $\chisq$ relevant to CP violation sensitivity, the test values of the CP violating phase \( \delta_{13} \) are fixed at $0$ or $\pi$. The  $\chisq$ is then computed for all possible true values of $( \delta_{13} )$ within the full range $[-\pi,\pi]$. 

In the analysis, marginalization is performed over all standard oscillation parameters (see Table~\ref{tab:parameters}) except for $\delta_{13}$ whose true value remains undetermined. Additionally, the total $\chisq$ is obtained by summing the contributions from two different oscillation channels: $(\nu_\mu \to \nu_e\,(\bar{\nu}_\mu \to \bar{\nu}_e))$ and $(\nu_\mu \to \nu_\mu \, (\bar{\nu}_\mu \to \bar{\nu}_\mu))$. Table~\ref{tab:sys} gives the detector configuration, efficiencies and systematic uncertainties used in our analysis.

\begin{table}[ht]
\arrayrulecolor{black}
\centering
\begin{tabular}{|l|ll|}
\hline
Detector details of DUNE  & \multicolumn{2}{c|}{Normalization error}  \\
                                              \cline{2-3}
                                              & \cellcolor{blue!5}Signal & \cellcolor{blue!5}Background                           \\
\hline
 $E_{p}:\,120$ GeV\,,  $L:\,1300$ km\, 
   &  & \\
\rowcolor{blue!5} target mass (fiducial) : 40 kt\, &   $\nu_e : 2\%$  & $\nu_e : 5\%$ \\ 
detector type : LArTPC\, & $\nu_\mu : 5\%$ & $\nu_\mu : 5\%$     \\ 
\rowcolor{blue!5}number of bins : 64\, &   & $\nu_\tau : 20\%$  \\
bin width : 0.125 GeV\,
&&  NC : 10\% \\
\cellcolor{blue!5} runtime (yr) $:\,6.5\, \nu + 6.5 \,\bar \nu\,$ &  &  \\
$R_\mu=0.20/{\sqrt E}$, $R_e=0.15/{\sqrt E}$   &   & \\
\hline
\end{tabular}
\vskip 0.1in
\caption{\label{tab:sys}
Detector configuration, efficiencies and systematic uncertainties for DUNE.}
\end{table}

{\bf{Sensitivity to the MH:}}  

The question of hierarchy of neutrino masses is a binary one i.e., the true hierarchy can be NH or IH. Here we address the following question. What is the sensitivity with which a particular experiment such as DUNE can distinguish between NH and IH. 
In order to understand the features of the sensitivity plots considering the true hierarchy as NH, we give a  statistical definition of $\chi^2$ as follows~\cite{Masud:2016nuj},
\bea
\chi^2_{} &\equiv&  \min_{\delta_{test}}  \sum_{i=1}^{x}  \sum_{j=\nu,\bar \nu }^{} 
 \frac{\left[N_{NH}^{i,j}(\delta_{true}) - 
 N_{IH}^{i,j} (\delta_{test}  )\right]^2 }
 {N_{NH}^{i,j} (\delta_{true})}\,.
 \label{eq:chisq_mh}
\eea \noindent
The quantities $N_{NH}^{i,j}$ and $N_{IH}^{i,j}$ denote the  number of events in the ${i,j}$-th bin for the NH and IH respectively.
The index $i$ labels the energy bins, where $i = 1$ to $x$, and $x$ is the total number of bins, specific to the experiment under consideration.  For DUNE experiment,
there are $x=62$ energy bins each having a width of $0.125$ GeV in the energy range of $0.5$ - $8$ GeV and $2$ bins of width 1 GeV each in the range $8$ - $10$ GeV~\cite{DUNE:2021cuw}. The index $j$ accounts for summing 
over the mode (neutrino or antineutrino). The  $\chi^2$ (defined in Eq.\eqref{eq:chisq_mh}) is computed for a given 
 set of true values by minimizing over the test parameters. This procedure is repeated for all possible
 true values specified in Table~\ref{tab:parameters}.

{\bf{Sensitivity to the octant of $\theta_{23}$:}}  

This pertains to distinguishing between the LO $(\theta_{23}<\pi/4)$ and the HO $(\theta_{23}>\pi/4)$.  The octant sensitivity at DUNE arises from both $\nu_{\mu}\to\nu_{e}$ and $\nu_{\mu}\to\nu_{\mu}$ channels~\cite{DUNE:2015lol}. To quantify the sensitivity to the octant of $\theta_{23}$, we employ the $\Delta \chi^2$ metric as
\bea
\Delta \chi^2_{\textrm{octant}} &=&
| \chi^2_{\theta_{23}^{test} > \pi/4} - \chi^2_{\theta_{23}^{test} < \pi/4} | \,,
\label{eq:chisq_oct}
\eea \noindent
where, the value of $\theta_{23}$ in the wrong octant is constrained to lie strictly within that octant and is not required to match the true value of $\sin^2 2\theta_{23}$.

  \section{Results and discussion}
  \label{sec:result}

\begin{figure}[t!]
\centering
\includegraphics[width=7in]{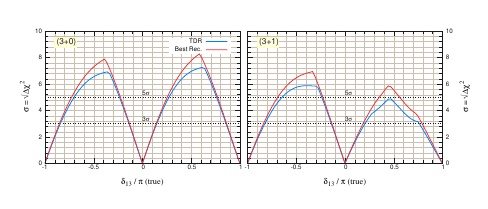}
\caption{\footnotesize{CP violation sensitivity as a function of $\delta_{13}$ for TDR (blue curve)  and best reconstruction case (red curve) at DUNE. The left panel depicts the $(3+0)$ case while the right panel depicts the 
$(3+1)$ case.}}
\label{fig:cp_sens}
\end{figure} \noindent
In this section, we present our results on DUNE sensitivities to CP violation, MH  and $\theta_{23}$ octant. 

 Fig.~\ref{fig:cp_sens} depicts the CP violation sensitivity as a function of true values of CP violating phase, $\delta_{13}$ for DUNE TDR (see blue curve) and best reconstruction case (see red curve). The left panel depicts the $(3+0)$ case while the right panel depicts the $(3+1)$ case.  The characteristic double peak shape implies that CP sensitivity is zero at the CP conserving values ($\delta_{13} = 0, \pi$), non-zero at other values ($\delta_{13} \neq 0, \pi$) and peaks at maximal CP violating values ($\delta_{13} = \pm \pi/2$). 
The best reconstruction case exhibits a  statistical advantage over DUNE TDR  both in $(3+0)$ and $(3+1)$ cases respectively. 

\begin{figure}[ht!]
\centering
\includegraphics[width=7in]{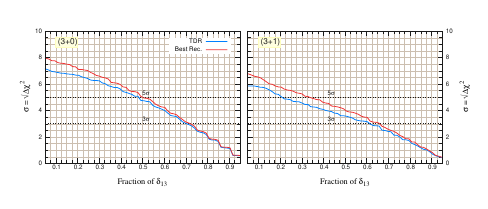}
\caption{\footnotesize{CP violation sensitivity as a function of fraction of $\delta_{13}$ for DUNE TDR (blue curve)  and best reconstruction (red curve). The left panel depicts the $(3+0)$ case while the right panel depicts the 
$(3+1)$ case.}}
\label{fig:frac_cp_sens}
\end{figure}\noindent
\begin{figure}[ht!]
\centering
\includegraphics[width=7in]{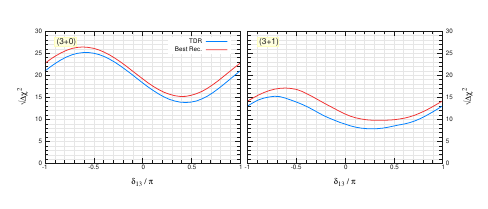}
\caption{\footnotesize{
MH sensitivity as a function of $\da$ for DUNE TDR (blue curve) and best reconstruction scenario (red curve). The left panel depicts the $(3+0)$ case while the right panel depicts the 
$(3+1)$ case.}}
\label{fig:mh_sens}  
\end{figure}\noindent
The CP violation sensitivity versus fraction of values of $\delta_{13}$ for DUNE TDR (blue curve) and the best reconstruction case (red curve) is shown in Fig.~\ref{fig:frac_cp_sens}.
 5$\sigma$ (3$\sigma$) sensitivity could be achieved by $\sim$ 48$\%$ (71$\%$) and $\sim$ 50$\%$ (73$\%$)  values of $\delta_{13}$ in $(3+0)$ case with DUNE TDR and best reconstruction scenario respectively. However, for the $(3+0)$ case, we have $\sim$ 22$\%$ (62$\%$) and $\sim$ 34$\%$ (67$\%$) values of $\delta_{13}$  with DUNE TDR and best reconstruction scenario respectively.

Fig.~\ref{fig:mh_sens} shows the MH sensitivity as a function of true values of $\da$ with DUNE TDR  and best reconstruction case. The MH sensitivity is already high in case of DUNE TDR for all values of the CP phase, $\delta_{13}$ in $(3+0)$ case. However, with best reconstruction scenario, sensitivity improves in comparison to DUNE TDR. Similar trend is seen in the $(3+1)$ case.
  
\begin{figure}[t!]
\centering   
\includegraphics[width=7in]{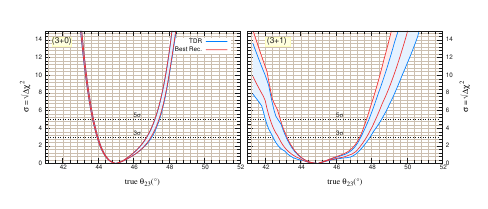}
\caption{\footnotesize{Sensitivity of octant of $\theta_{23}$ as a function of $\theta_{23}$ for DUNE TDR (blue band)  and best reconstruction case (pink band). The left panel depicts the $(3+0)$ case while the right panel depicts the 
$(3+1)$ case.}}
\label{fig:oct_sens}
\end{figure}

Fig.~\ref{fig:oct_sens} shows the sensitivity of determining the octant  of $\theta_{23}$ as a function of the true values of $\theta_{23}$ for DUNE TDR ({blue} band) and best reconstruction case ({{pink band}}).  
The octant sensitivity is  better for $\theta_{23}$ lying in the LO for the $(3+0)$ case. The $(3+1)$ case worsens the octant determination but is better for $\theta_{23}$ lying in the HO.
%

 \noindent

\section{Conclusion}
\label{sec:conclusion}

The key physics goals of the long baseline experiments such as DUNE are to resolve if CP is violated in the leptonic sector, to establish whether the MH is normal or inverted and to figure out what the correct octant of $\theta_{23}$ is. In the standard configuration,  DUNE would have a total runtime of 13 years (distributed equally in the $\nu$ and $\bar \nu$ mode) with the standard LE beam.  The LE beam that is often used in DUNE simulations has a peak around $2$ - $3$ GeV  but very sharply falls off at $E \gsim 4$ GeV. In an earlier work involving some of the authors, we have obtained optimal configuration of runtime and different beam tunes which allow for attaining best sensitivities to these standard unknowns~\cite{Rout:2020cxi}. It has been conclusively shown in literature that  any BSM physics scenario such as sterile neutrinos or NSI leads to spoiling of sensitivities of these unknowns~\cite{Gandhi:2015xza,Dutta:2016glq,Masud:2015xva,deGouvea:2015ndi,Coloma:2015kiu,Masud:2016nuj,Agarwalla:2016xlg,Agarwalla:2018nlx,Capozzi:2019iqn,Chatterjee:2020kkm,Chatterjee:2022pqg}.

It is then interesting to explore the role played by energy resolution and how it impacts DUNE sensitivities in presence of   BSM physics. The impact of energy resolution on a  particular BSM physics scenario of neutrino NSI has been explored in~\cite{Chatterjee:2021wac}. In the present work, we have studied the impact of improved energy resolution on DUNE sensitivity to a particular BSM physics scenario of an additional light sterile neutrino, i.e., $(3+1)$ case and compare with results for the standard $(3+0)$ case.
 Better energy resolution would also aid in disentangling parameter degeneracies among effects arising from standard three neutrino oscillations ($(3+0)$ case) and BSM physics scenario ($(3+1)$ case). We consider two cases : standard DUNE TDR and best resolution scenario. As expected using better energy resolution leads to   improvement in attaining the key physics goals.
 
Let us first examine the CP violation sensitivity as depicted in Fig.~\ref{fig:cp_sens}-\ref{fig:frac_cp_sens} and Table~\ref{tab:phys}. We conclude that with a fixed exposure of 624 kt$\cdot$MW$\cdot$yr and nominal energy resolution as given in TDR, DUNE will be able to reach $5\,\sigma$ ($3\,\sigma$) discovery of CP violation for $48\,\%$ ($71\,\%$) of $\da$ values for $(3+0)$ case and $22\,\%$ ($62\,\%$) of $\da$ values for $(3+1)$ case. However, there is improvement with
better energy resolution.

The same can be inferred regarding MH (see Fig~\ref{fig:mh_sens}) i.e., with better energy resolution we have improved sensitivity to neutrino MH.

The octant sensitivity is better for $\theta_{23}$ lying in the LO for the $(3+0)$ case. The $(3+1)$ case worsens the octant determination but is better for $\theta_{23}$ lying in the HO (see Fig~\ref{fig:oct_sens}).

\begin{table}[h]
\arrayrulecolor{black}
\centering

\scalebox{0.9}{
\begin{tabular}{|c|c|c|c|c|}
\hline
\multirow{2}{*}{Physics goal : CP violation} & 
\multicolumn{2}{c|}{\cellcolor{blue!5}$\%$ of $\delta_{13}$ values for $(3+0)$ case} & %
   \multicolumn{2}{c|}{\cellcolor{blue!5}$\%$ of $\delta_{13}$ values for $(3+1)$ case} \\
\cline{2-5}
 &  ~~~~~~~TDR~~~~~~~ &  Best Rec. &  ~~~~~~~TDR~~~~~~~ &  Best Rec.  \\
\hline
 3$\sigma$    & \cellcolor{blue!5}71 & \cellcolor{blue!5}73 & \cellcolor{blue!5}62 & \cellcolor{blue!5}67 \\
\hline
5$\sigma$ & 48 & 50 & 22 & 34 \\
\hline
\end{tabular}}
\vskip 0.1in
\caption{\footnotesize{\label{tab:phys}
Percentage of $\delta_{13}$ values for which DUNE will achieve 3$\sigma$ or 5$\sigma$ CP violation sensitivity for the fixed exposure considered in the present study.}}
\end{table} 

\section*{Acknowledgments}
SP acknowledges JNU for support in the form of fellowship. The  numerical analysis has been performed using the HPC cluster at SPS, JNU funded by DST-FIST. JR would like to thank OSHEC (Odisha State
Higher Education Council) for the financial support through Mukhyamantri Research Innovation (MRI) for Extramural Research Funding 2023. 
The research of PM is supported in part by the International Centre for Theoretical Sciences (ICTS) for participating in the discussion meeting - (QM100) A Hundred Years of Quantum Mechanics (code: ICTS/qm100-2025/01)
and 
by the Inter-University Centre for Astronomy and Astrophysics (IUCAA), Pune through its Associateship Programme.
\\
 This work reflects the views of the authors and not those of the DUNE collaboration.

\appendix
\renewcommand{\theequation}{\thesection.\arabic{equation}}
\setcounter{equation}{0}
\counterwithin{figure}{section}
\numberwithin{equation}{section}
\renewcommand{\thesection}{\Alph{section}}
\renewcommand{\thesubsection}{\Alph{subsection}}
\section{CP and MH probability differences for the $(3+0)$ case}
 \label{app_1}
For the $(3+0)$ scenario, we use the expressions derived in~\cite{Akhmedov:2004ny}. In this section, we present the approximate expressions of CP and MH probability differences for $\nu_{\mu} \to \nu_{e}$ and
 $\nu_{\mu} \to \nu_{\mu}$ channels. We adopt the parameterization of the mixing matrix as given in Eq.~\eqref{eq:PMNS}.

The CP violating probability differences for $\nu_{\mu} \to \nu_{e}$ channel, i.e., $\Delta P^{{CP} (3+0)}_{\mu e}$ is given by
\bea
\Delta P^{{CP} (3+0)}_{\mu e} &\simeq& 4 s_{13}^2  s_{23}^2  \Bigg[\frac{\sin^{2}(\hat A-1) \Delta}{(\hat A-1)^{2}} -  \frac{\sin^{2}(\hat A+1) \Delta}{(\hat A+1)^{2}} \Bigg] - 8\alpha
  s_{13} s_{12} 
         s_{23} c_{12} 
        c_{23}\sin \delta_{13} \nonumber \\ &&
 \frac{\sin \hat A \Delta}{\hat A} \Big[\cos \Delta \cot \delta_{13} \Theta_{-} + \sin \Delta \Theta_{+} \Big]+ \mathcal{O}(\lambda^{4})\,,  
\label{eq:Delta_CP_mue}
\eea
The CP violating probability differences for $\nu_{\mu} \to \nu_{\mu}$ channel, i.e., $\Delta P^{{CP} (3+0)}_{\mu \mu}$ is given by
\bea
\Delta P^{{CP} (3+0)}_{\mu\mu} &\simeq& \frac{1}{2\hat A} \alpha^2 \sin^2 2\theta_{12}  \sin^22\theta_{23} \Bigg[\frac{1}{2\hat A} \sin2 \hat A \Delta \sin 2 \Delta - \Delta \sin 2 \Delta \Bigg] \ \nonumber\\
   &+& 4 s_{13}^2 s_{23}^2 \Bigg[\frac{\sin^{2}(\hat A+1) \Delta}{(\hat A+1)^{2}} -  \frac{\sin^{2}(\hat A-1) \Delta}{(\hat A-1)^{2}} \Bigg]- 2 s_{13}^2 \sin^22\theta_{23}\Bigg\{\frac{\hat A \Delta}{2}
    \Bigg( \frac{\sin 2\Delta}{\hat A+1} - \frac{\sin 2\Delta}{\hat A-1}\Bigg)  \nonumber\\
    &+& \sin \Delta \cos \hat A \Delta \left(\frac{\sin(\hat A+1) \Delta}{(\hat A+1)^{2}} + \frac{\sin(\hat A-1) \Delta}{(\hat A-1)^{2}} \right) \Bigg \} \nonumber \\
   &+& 2 \alpha s_{13}  \sin 2\theta_{12}  \sin 2\theta_{23}
       \cos\delta_{13} \cos\Delta \frac{\sin \hat A\Delta}{\hat A} 
  \Theta_{-} \nonumber\\
   &+& 2\alpha s_{13} \sin 2\theta_{12}  
       \sin2\theta_{23} \cos2\theta_{23} \cos\delta_{13} 
   \sin\Delta \Bigg\{\hat A \Bigg(
 \frac{\sin \Delta}{\hat A-1} - \frac{\sin \Delta}{\hat A+1} \Bigg) \, \nonumber \\
 &-& \frac{\sin \hat A \Delta}{\hat A} \Bigg(\frac{\cos(\hat A+1) \Delta}{\hat A+1} + \frac{\cos(\hat A-1) \Delta}{\hat A-1} \Bigg) \Bigg\}+ \mathcal{O}(\lambda^{4}) \,,
\label{eq:Delta_CP_mm} 
       \eea
The MH probability differences for $\nu_{\mu} \to \nu_{e}$ channel, i.e., $\Delta P^{{MH} (3+0)}_{\mu e}$ is given by
       \bea
     \Delta P^{{MH}(3+0)}_{\mu e} &\simeq&   4 s_{13}^2  s_{23}^2  \left[\dfrac{\sin^{2}[\hat A-1] \Delta}{(\hat A-1)^{2}} -  \frac{\sin^{2}[\hat A+1] \Delta}{(\hat A+1)^{2}} \right]+ 2  \alpha  s_{13} \sin 2\theta_{12} 
 \sin 2 \theta_{23} \dfrac{\sin \hat A \Delta}{\hat A}\sin \delta_{13} \nonumber \\ &&  \Big[\Theta_{+}\cos \Delta \cot \delta_{13} + \Theta_{-} \sin \Delta \Big]+ \mathcal{O}(\lambda^{4})\,,
 \label{eq:Delta_MH_me}
  \eea
The MH violating probability differences for $\nu_{\mu} \to \nu_{\mu}$ channel, i.e., $\Delta P^{{MH} (3+0)}_{\mu \mu}$ is given by
    \bea
    \Delta P^{{MH}(3+0)}_{\mu\mu} &\simeq& 
    2 \alpha c^{2}_{12} \sin^22\theta_{23} \Delta \sin 2 \Delta +  
    \frac{1}{2\hat A} \alpha^2 \sin^2 2\theta_{12}  \sin^2 2\theta_{23}     \Bigg\{\frac{\Delta \sin 2 \Delta}{\hat A}\Big[\cos(\hat{A}-1)\Delta \nonumber \\ &+& \cos(\hat{A}+1)\Delta\Big]-\Delta \sin \Delta\Bigg\}+ 4 s_{13}^2 s_{23}^2 \left[\frac{\sin^{2}(\hat A+1) \Delta}{(\hat A+1)^{2}} -  \frac{\sin^{2}(\hat A-1) \Delta}{(\hat A-1)^{2}} \right]   \nonumber\\ &-& 2 s_{13}^2 \sin^22\theta_{23}\Bigg\{\frac{\hat A \Delta}{2}
    \Bigg[\frac{\sin 2\Delta}{\hat A+1} + \frac{\sin 2\Delta}{\hat A-1}\Bigg]+ \sin \Delta \cos \hat A \Delta \left[\frac{\sin(\hat A+1) \Delta}{(\hat A+1)^{2}} + \frac{\sin(\hat A-1) \Delta}{(\hat A-1)^{2}} \right] \Bigg \} \nonumber \\
   &-& 2 \alpha s_{13}  \sin 2\theta_{12}  \sin 2\theta_{23}
       \cos\delta_{13} \cos\Delta \frac{\sin \hat A\Delta}{\hat A} 
  \Theta_{+} + 2\alpha s_{13} \sin 2\theta_{12}  
       \sin2\theta_{23} \cos2\theta_{23} \cos\delta_{13} 
   \sin\Delta \nonumber \\ && \Bigg\{\hat A \left[
 \frac{\sin \Delta}{\hat A-1} + \frac{\sin \Delta}{\hat A+1} \right]
 + \frac{\sin \hat A \Delta}{\hat A} \left[\frac{\cos(\hat A+1) \Delta}{\hat A+1} - \frac{\cos(\hat A-1) \Delta}{\hat A-1} \right] \Bigg\} + \mathcal{O}(\lambda^{4})\,,
\label{eq:Delta_MH_mm}
\eea
       where ,
       \bea
       \Theta_{\pm} = \left[\dfrac{\sin(\hat A+1) \Delta}{\hat A+1} \pm \dfrac{\sin(\hat A-1) \Delta}{\hat A-1} \right] \,.\nonumber 
       \eea
\bibliographystyle{unsrt}
\bibliography{reference}

\begin{thebibliography}{10}

\bibitem{nobel2015}
T.~Kajita and A.~B. McDonald.
\newblock For the discovery of neutrino oscillations, which shows that
  neutrinos have mass.
\newblock The Nobel Prize in Physics
  2015.\url{https://www.nobelprize.org/prizes/physics/2015/summary/}.

\bibitem{deSalas:2020pgw}
P.~F. de~Salas, D.~V. Forero, S.~Gariazzo, P.~Mart\'\i{}nez-Mirav\'e, O.~Mena,
  C.~A. Ternes, M.~T\'ortola, and J.~W.~F. Valle.
\newblock {2020 global reassessment of the neutrino oscillation picture}.
\newblock {\em JHEP}, 02:071, 2021.

\bibitem{Capozzi:2021fjo}
Francesco Capozzi, Eleonora Di~Valentino, Eligio Lisi, Antonio Marrone,
  Alessandro Melchiorri, and Antonio Palazzo.
\newblock {Unfinished fabric of the three neutrino paradigm}.
\newblock {\em Phys. Rev. D}, 104(8):083031, 2021.

\bibitem{Esteban:2024eli}
Ivan Esteban, M.~C. Gonzalez-Garcia, Michele Maltoni, Ivan Martinez-Soler,
  Jo\~ao~Paulo Pinheiro, and Thomas Schwetz.
\newblock {NuFit-6.0: Updated global analysis of three-flavor neutrino
  oscillations}.
\newblock 10 2024.

\bibitem{Davidson:2008bu}
Sacha Davidson, Enrico Nardi, and Yosef Nir.
\newblock {Leptogenesis}.
\newblock {\em Phys. Rept.}, 466:105--177, 2008.

\bibitem{Giunti:2019aiy}
Carlo Giunti and T.~Lasserre.
\newblock {eV-scale Sterile Neutrinos}.
\newblock {\em Ann. Rev. Nucl. Part. Sci.}, 69:163--190, 2019.

\bibitem{Dasgupta:2021ies}
Basudeb Dasgupta and Joachim Kopp.
\newblock {Sterile Neutrinos}.
\newblock {\em Phys. Rept.}, 928:1--63, 2021.

\bibitem{LSND:1996ubh}
C.~Athanassopoulos et~al.
\newblock {Evidence for anti-muon-neutrino ---\ensuremath{>}
  anti-electron-neutrino oscillations from the LSND experiment at LAMPF}.
\newblock {\em Phys. Rev. Lett.}, 77:3082--3085, 1996.

\bibitem{MiniBooNE:2020pnu}
A.~A. Aguilar-Arevalo et~al.
\newblock {Updated MiniBooNE neutrino oscillation results with increased data
  and new background studies}.
\newblock {\em Phys. Rev. D}, 103(5):052002, 2021.

\bibitem{MicroBooNE:2022sdp}
P.~Abratenko et~al.
\newblock {First Constraints on Light Sterile Neutrino Oscillations from
  Combined Appearance and Disappearance Searches with the MicroBooNE Detector}.
\newblock {\em Phys. Rev. Lett.}, 130(1):011801, 2023.

\bibitem{MiniBooNE:2022emn}
A.~A. Aguilar-Arevalo et~al.
\newblock {MiniBooNE and MicroBooNE Combined Fit to a 3+1 Sterile Neutrino
  Scenario}.
\newblock {\em Phys. Rev. Lett.}, 129(20):201801, 2022.

\bibitem{SAGE:2009eeu}
J.~N. Abdurashitov et~al.
\newblock {Measurement of the solar neutrino capture rate with gallium metal.
  III: Results for the 2002--2007 data-taking period}.
\newblock {\em Phys. Rev. C}, 80:015807, 2009.

\bibitem{Kaether:2010ag}
F.~Kaether, W.~Hampel, G.~Heusser, J.~Kiko, and T.~Kirsten.
\newblock {Reanalysis of the GALLEX solar neutrino flux and source
  experiments}.
\newblock {\em Phys. Lett. B}, 685:47--54, 2010.

\bibitem{Barinov:2021asz}
V.~V. Barinov et~al.
\newblock {Results from the Baksan Experiment on Sterile Transitions (BEST)}.
\newblock {\em Phys. Rev. Lett.}, 128(23):232501, 2022.

\bibitem{Huber:2011wv}
Patrick Huber.
\newblock {On the determination of anti-neutrino spectra from nuclear
  reactors}.
\newblock {\em Phys. Rev. C}, 84:024617, 2011.
\newblock [Erratum: Phys.Rev.C 85, 029901 (2012)].

\bibitem{Mention:2011rk}
G.~Mention, M.~Fechner, Th. Lasserre, Th.~A. Mueller, D.~Lhuillier, M.~Cribier,
  and A.~Letourneau.
\newblock {The Reactor Antineutrino Anomaly}.
\newblock {\em Phys. Rev. D}, 83:073006, 2011.

\bibitem{Serebrov:2020kmd}
A.~P. Serebrov et~al.
\newblock {Search for sterile neutrinos with the Neutrino-4 experiment and
  measurement results}.
\newblock {\em Phys. Rev. D}, 104(3):032003, 2021.

\bibitem{Machado:2019oxb}
Pedro~AN Machado, Ornella Palamara, and David~W Schmitz.
\newblock {The Short-Baseline Neutrino Program at Fermilab}.
\newblock {\em Ann. Rev. Nucl. Part. Sci.}, 69:363--387, 2019.

\bibitem{SBND:2024vgn}
P.~Abratenko et~al.
\newblock {Scintillation light in SBND: simulation, reconstruction, and
  expected performance of the photon detection system}.
\newblock {\em Eur. Phys. J. C}, 84(10):1046, 2024.

\bibitem{Torretta:2024fbn}
D.~Torretta.
\newblock {Status and perspectives of the ICARUS experiment at the Fermilab
  Short Baseline Neutrino program}.
\newblock {\em JINST}, 19(04):C04061, 2024.

\bibitem{Hyper-Kamiokande:2018ofw}
K.~Abe et~al.
\newblock {Hyper-Kamiokande Design Report}.
\newblock {\em arXiv:1805.04163}.

\bibitem{DUNE:2020ypp}
Babak Abi et~al.
\newblock {Deep Underground Neutrino Experiment (DUNE), Far Detector Technical
  Design Report, Volume II: DUNE Physics}.
\newblock 2 2020.

\bibitem{DUNE:2021cuw}
B.~Abi et~al.
\newblock {Experiment Simulation Configurations Approximating DUNE TDR}.
\newblock {\em arXiv:2103.04797}.

\bibitem{Akindinov:2019flp}
A.~V. Akindinov et~al.
\newblock {Letter of Interest for a Neutrino Beam from Protvino to
  KM3NeT/ORCA}.
\newblock {\em Eur. Phys. J. C}, 79(9):758, 2019.

\bibitem{KM3NET:2016zxf}
S.~Adrian-Martinez et~al.
\newblock {Letter of intent for KM3NeT 2.0}.
\newblock {\em J. Phys. G}, 43(8):084001, 2016.

\bibitem{Goswami:1995yq}
Srubabati Goswami.
\newblock {Accelerator, reactor, solar and atmospheric neutrino oscillation:
  Beyond three generations}.
\newblock {\em Phys. Rev. D}, 55:2931--2949, 1997.

\bibitem{Gandhi:2015xza}
Raj Gandhi, Boris Kayser, Mehedi Masud, and Suprabh Prakash.
\newblock {The impact of sterile neutrinos on CP measurements at long
  baselines}.
\newblock {\em JHEP}, 11:039, 2015.

\bibitem{Parke:2015goa}
Stephen Parke and Mark Ross-Lonergan.
\newblock {Unitarity and the three flavor neutrino mixing matrix}.
\newblock {\em Phys. Rev. D}, 93(11):113009, 2016.

\bibitem{Dutta:2016glq}
Debajyoti Dutta, Raj Gandhi, Boris Kayser, Mehedi Masud, and Suprabh Prakash.
\newblock {Capabilities of long-baseline experiments in the presence of a
  sterile neutrino}.
\newblock {\em JHEP}, 11:122, 2016.

\bibitem{Gandhi:2017vzo}
Raj Gandhi, Boris Kayser, Suprabh Prakash, and Samiran Roy.
\newblock {What measurements of neutrino neutral current events can reveal}.
\newblock {\em JHEP}, 11:202, 2017.

\bibitem{Kosmas:2017zbh}
T.~S. Kosmas, D.~K. Papoulias, M.~Tortola, and J.~W.~F. Valle.
\newblock {Probing light sterile neutrino signatures at reactor and Spallation
  Neutron Source neutrino experiments}.
\newblock {\em Phys. Rev. D}, 96(6):063013, 2017.

\bibitem{Choubey:2017cba}
Sandhya Choubey, Debajyoti Dutta, and Dipyaman Pramanik.
\newblock {Imprints of a light Sterile Neutrino at DUNE, T2HK and T2HKK}.
\newblock {\em Phys. Rev. D}, 96(5):056026, 2017.

\bibitem{Agarwalla:2018nlx}
Sanjib~Kumar Agarwalla, Sabya~Sachi Chatterjee, and Antonio Palazzo.
\newblock {Signatures of a Light Sterile Neutrino in T2HK}.
\newblock {\em JHEP}, 04:091, 2018.

\bibitem{Reyimuaji:2019wbn}
Yakefu Reyimuaji and Chun Liu.
\newblock {Prospects of light sterile neutrino searches in long-baseline
  neutrino oscillations}.
\newblock {\em JHEP}, 06:094, 2020.

\bibitem{Chatterjee:2022pqg}
Animesh Chatterjee, Srubabati Goswami, and Supriya Pan.
\newblock {Matter effect in presence of a sterile neutrino and resolution of
  the octant degeneracy using a liquid argon detector}.
\newblock {\em Phys. Rev. D}, 108(9):095050, 2023.

\bibitem{Singha:2022btw}
Dinesh~Kumar Singha, Monojit Ghosh, Rudra Majhi, and Rukmani Mohanta.
\newblock {Study of light sterile neutrino at the long-baseline experiment
  options at KM3NeT}.
\newblock {\em Phys. Rev. D}, 107(7):075039, 2023.

\bibitem{Chattopadhyay:2022hkw}
Dibya~S. Chattopadhyay, Moon~Moon Devi, Amol Dighe, Debajyoti Dutta, Dipyaman
  Pramanik, and Sushant~K. Raut.
\newblock {Sterile neutrinos: propagation in matter and sensitivity to sterile
  mass ordering}.
\newblock {\em JHEP}, 02:044, 2023.

\bibitem{Majhi:2019hdj}
Rudra Majhi, C.~Soumya, and Rukmani Mohanta.
\newblock {Light sterile neutrinos and their implications on currently running
  long-baseline and neutrinoless double beta decay experiments}.
\newblock {\em J. Phys. G}, 47(9):095002, 2020.

\bibitem{Fiza:2021gvq}
Nishat Fiza, Mehedi Masud, and Manimala Mitra.
\newblock {Exploring the new physics phases in 3+1 scenario in neutrino
  oscillation experiments}.
\newblock {\em JHEP}, 09:162, 2021.

\bibitem{Parveen:2023ixk}
Sabila Parveen, Kiran Sharma, Sudhanwa Patra, and Poonam Mehta.
\newblock {Signals of eV-scale sterile neutrino at long baseline neutrino
  experiments}.
\newblock {\em Eur. Phys. J. C}, 85(2):181, 2025.

\bibitem{Parveen:2024bcc}
Sabila Parveen, Mehedi Masud, Mary Bishai, and Poonam Mehta.
\newblock {Sterile sector impacting the correlations and degeneracies among
  mixing parameters at the Deep Underground Neutrino Experiment}.
\newblock {\em JHEP}, 01:139, 2025.

\bibitem{DeRomeri:2016qwo}
Valentina De~Romeri, Enrique Fernandez-Martinez, and Michel Sorel.
\newblock {Neutrino oscillations at DUNE with improved energy reconstruction}.
\newblock {\em JHEP}, 09:030, 2016.

\bibitem{Chatterjee:2021wac}
Sabya~Sachi Chatterjee, P.~S.~Bhupal Dev, and Pedro A.~N. Machado.
\newblock {Impact of improved energy resolution on DUNE sensitivity to neutrino
  non-standard interactions}.
\newblock {\em JHEP}, 08:163, 2021.

\bibitem{Pontecorvo:1957qd}
B.~Pontecorvo.
\newblock {Inverse beta processes and nonconservation of lepton charge}.
\newblock {\em Zh. Eksp. Teor. Fiz.}, 34:247, 1957.

\bibitem{Pontecorvo:1957cp}
B.~Pontecorvo.
\newblock {Mesonium and anti-mesonium}.
\newblock {\em Sov. Phys. JETP}, 6:429, 1957.

\bibitem{Maki:1962mu}
Ziro Maki, Masami Nakagawa, and Shoichi Sakata.
\newblock {Remarks on the unified model of elementary particles}.
\newblock {\em Prog. Theor. Phys.}, 28:870--880, 1962.

\bibitem{Gribov:1968kq}
V.N. Gribov and B.~Pontecorvo.
\newblock {Neutrino astronomy and lepton charge}.
\newblock {\em Phys. Lett. B}, 28:493, 1969.

\bibitem{ParticleDataGroup:2024cfk}
S.~Navas et~al.
\newblock {Review of particle physics}.
\newblock {\em Phys. Rev. D}, 110(3):030001, 2024.

\bibitem{Kopp:2013vaa}
Joachim Kopp, Pedro A.~N. Machado, Michele Maltoni, and Thomas Schwetz.
\newblock {Sterile Neutrino Oscillations: The Global Picture}.
\newblock {\em JHEP}, 05:050, 2013.

\bibitem{Klop:2014ima}
N.~Klop and A.~Palazzo.
\newblock {Imprints of CP violation induced by sterile neutrinos in T2K data}.
\newblock {\em Phys. Rev. D}, 91(7):073017, 2015.

\bibitem{DUNE:2020fgq}
B.~Abi et~al.
\newblock {Prospects for beyond the Standard Model physics searches at the Deep
  Underground Neutrino Experiment}.
\newblock {\em Eur. Phys. J. C}, 81(4):322, 2021.

\bibitem{Dentler:2018sju}
Mona Dentler, \'Alvaro Hern\'andez-Cabezudo, Joachim Kopp, Pedro A.~N. Machado,
  Michele Maltoni, Ivan Martinez-Soler, and Thomas Schwetz.
\newblock {Updated Global Analysis of Neutrino Oscillations in the Presence of
  eV-Scale Sterile Neutrinos}.
\newblock {\em JHEP}, 08:010, 2018.

\bibitem{Akhmedov:2004ny}
Evgeny~K. Akhmedov, Robert Johansson, Manfred Lindner, Tommy Ohlsson, and
  Thomas Schwetz.
\newblock {Series expansions for three flavor neutrino oscillation
  probabilities in matter}.
\newblock {\em JHEP}, 04:078, 2004.

\bibitem{Huber:2004ka}
Patrick Huber, M.~Lindner, and W.~Winter.
\newblock {Simulation of long-baseline neutrino oscillation experiments with
  GLoBES (General Long Baseline Experiment Simulator)}.
\newblock {\em Comput. Phys. Commun.}, 167:195, 2005.

\bibitem{Huber:2007ji}
Patrick Huber, Joachim Kopp, Manfred Lindner, Mark Rolinec, and Walter Winter.
\newblock {New features in the simulation of neutrino oscillation experiments
  with GLoBES 3.0: General Long Baseline Experiment Simulator}.
\newblock {\em Comput. Phys. Commun.}, 177:432--438, 2007.

\bibitem{Dziewonski:1981xy}
A.~M. Dziewonski and D.~L. Anderson.
\newblock {Preliminary reference earth model}.
\newblock {\em Phys. Earth Planet. Interiors}, 25:297--356, 1981.

\bibitem{Agostinelli:2002hh}
S.~Agostinelli et~al.
\newblock {GEANT4: A Simulation toolkit}.
\newblock {\em Nucl. Instrum. Meth.}, A506:250--303, 2003.

\bibitem{Allison:2006ve}
John Allison et~al.
\newblock {Geant4 developments and applications}.
\newblock {\em IEEE Trans. Nucl. Sci.}, 53:270, 2006.

\bibitem{Gandhi:2004md}
Raj Gandhi, Pomita Ghoshal, Srubabati Goswami, Poonam Mehta, and S.~Uma Sankar.
\newblock {Large matter effects in nu(mu) ---\ensuremath{>} nu(tau)
  oscillations}.
\newblock {\em Phys. Rev. Lett.}, 94:051801, 2005.

\bibitem{Gandhi:2004bj}
Raj Gandhi, Pomita Ghoshal, Srubabati Goswami, Poonam Mehta, and S.~Uma Sankar.
\newblock {Earth matter effects at very long baselines and the neutrino mass
  hierarchy}.
\newblock {\em Phys. Rev. D}, 73:053001, 2006.

\bibitem{Kelly:2018kmb}
Kevin~J. Kelly and Stephen~J. Parke.
\newblock {Matter Density Profile Shape Effects at DUNE}.
\newblock {\em Phys. Rev. D}, 98(1):015025, 2018.

\bibitem{Chatterjee:2018dyd}
Animesh Chatterjee, Felipe Kamiya, Celio~A. Moura, and Jaehoon Yu.
\newblock {Impact of Matter Density Profile Shape on Non-Standard Interactions
  at DUNE}.
\newblock {\em arXiv:1809.09313}.

\bibitem{Abi:2020loh}
B.~Abi et~al.
\newblock Deep underground neutrino experiment (dune), far detector technical
  design report, volume i: Introduction to dune.
\newblock {\em JINST}, 15:T08008, 2020.

\bibitem{Friedland:2018vry}
Alexander Friedland and Shirley~Weishi Li.
\newblock {Understanding the energy resolution of liquid argon neutrino
  detectors}.
\newblock {\em Phys. Rev. D}, 99(3):036009, 2019.

\bibitem{Bass:2013vcg}
M.~Bass et~al.
\newblock {Baseline Optimization for the Measurement of CP Violation, Mass
  Hierarchy, and $\theta_{23}$ Octant in a Long-Baseline Neutrino Oscillation
  Experiment}.
\newblock {\em Phys. Rev. D}, 91(5):052015, 2015.

\bibitem{Masud:2016nuj}
Mehedi Masud and Poonam Mehta.
\newblock {Nonstandard interactions and resolving the ordering of neutrino
  masses at DUNE and other long baseline experiments}.
\newblock {\em Phys. Rev. D}, 94(5):053007, 2016.

\bibitem{Messier:1999kj}
Mark~D. Messier.
\newblock {\em {Evidence for neutrino mass from observations of atmospheric
  neutrinos with Super-Kamiokande}}.
\newblock PhD thesis, Boston U., 1999.

\bibitem{Masud:2015xva}
Mehedi Masud, Animesh Chatterjee, and Poonam Mehta.
\newblock {Probing CP violation signal at DUNE in presence of non-standard
  neutrino interactions}.
\newblock {\em J. Phys. G}, 43(9):095005, 2016.

\bibitem{Rout:2020emr}
Jogesh Rout, Sheeba Shafaq, Mary Bishai, and Poonam Mehta.
\newblock {Physics prospects with the second oscillation maximum at the Deep
  Underground Neutrino Experiment}.
\newblock {\em Phys. Rev. D}, 103(11):116003, 2021.

\bibitem{Rout:2020cxi}
Jogesh Rout, Samiran Roy, Mehedi Masud, Mary Bishai, and Poonam Mehta.
\newblock {Impact of high energy beam tunes on the sensitivities to the
  standard unknowns at DUNE}.
\newblock {\em Phys. Rev. D}, 102:116018, 2020.

\bibitem{Qian:2012zn}
X.~Qian, A.~Tan, W.~Wang, J.~J. Ling, R.~D. McKeown, and C.~Zhang.
\newblock {Statistical Evaluation of Experimental Determinations of Neutrino
  Mass Hierarchy}.
\newblock {\em Phys. Rev.}, D86:113011, 2012.

\bibitem{DUNE:2015lol}
R.~Acciarri et~al.
\newblock {Long-Baseline Neutrino Facility (LBNF) and Deep Underground Neutrino
  Experiment (DUNE)}: {Conceptual Design Report, Volume 2: The Physics Program
  for DUNE at LBNF}.
\newblock 12 2015.

\bibitem{deGouvea:2015ndi}
Andr\'e de~Gouv\^ea and Kevin~J. Kelly.
\newblock {Non-standard neutrino interactions at DUNE}.
\newblock {\em Nucl. Phys. B}, 908:318--335, 2016.

\bibitem{Coloma:2015kiu}
Pilar Coloma.
\newblock {Non-Standard Interactions in propagation at the Deep Underground
  Neutrino Experiment}.
\newblock {\em JHEP}, 03:016, 2016.

\bibitem{Agarwalla:2016xlg}
Sanjib~Kumar Agarwalla, Sabya~Sachi Chatterjee, and Antonio Palazzo.
\newblock {Octant of $\theta_{23}$ in danger with a light sterile neutrino}.
\newblock {\em Phys. Rev. Lett.}, 118(3):031804, 2017.

\bibitem{Capozzi:2019iqn}
Francesco Capozzi, Sabya~Sachi Chatterjee, and Antonio Palazzo.
\newblock {Neutrino Mass Ordering Obscured by Nonstandard Interactions}.
\newblock {\em Phys. Rev. Lett.}, 124(11):111801, 2020.

\bibitem{Chatterjee:2020kkm}
Sabya~Sachi Chatterjee and Antonio Palazzo.
\newblock {Nonstandard Neutrino Interactions as a Solution to the $NO\nu A$ and
  T2K Discrepancy}.
\newblock {\em Phys. Rev. Lett.}, 126(5):051802, 2021.

\end{thebibliography}
\end{document}